\documentclass[a4paper,11pt]{article}
\usepackage{latexsym,amssymb}
\begin{document}
\begin{center}{\Large {\bf BTW model with probabilistically nonuniform 
distribution of particles from the unstable sites}}\end{center}

\vskip 1cm

\begin{center}{\it  Ajanta Bhowal Acharyya}\\
{\it Department of Physics, Lady Brabourne College}\\
{\it P-1/2 Suhrawardy Avenue, Calcutta - 700017, India}\\
{ E-mail:ajanta.bhowal@gmail.com}\end{center}

\noindent 
{\bf Abstract:} The two dimensional BTW model of SOC, with probabilistically nonuniform
distribution of particles among the nearest neighbouring sites, is studied by 
computer simulation.  When the value of height variable of a particular site 
reaches the critical value ($z_c=4$), the value of height variable of that site 
is reduced by four units by distributing four particles among the four nearest
neighbouring sites. In this paper, it is considered that, two particles are 
distributed equally among the two nearest neighbouring sites along X-axis.
However, the distribution of other two particles along Y-axis, is 
probabilistically nonuniform.  The variation of spatial average of the height 
variable with time is studied.  In the SOC state, the distribution of avalanche
sizes and durations is obtained.  The  total number of topplings occured during
the stipulated time of evolution is also calculated.

\vspace {1 cm}
\leftline {\bf Keywords: BTW model, Height variable, Avalanche size, SOC}
\leftline {\bf PACS Numbers: 05.50 +q}

\vspace {3cm}
\newpage

\leftline {\bf I. Introduction}

There exists some extended driven systems in nature, like forest fire,
earthquake, growth of sandpile, which can be explained by using  self-organised
criticality (SOC).  This phenomena of SOC is characterised by sponteneous
evolution into a steady state which shows long-range spatial and temporal
correlations. The concept of SOC was introduced by Bak, Tang and  Wiesenfeld
in terms of a simple cellular automata model\cite{btw1,btw2}.  The steady state
dynamics of the model shows a power law behaviour in the probability
distributions for the occurence of the relaxation (avalanches) clusters of a
certain size, area, lifetime, etc. Extensive work has been done so far to
study the properties of the model in the steady SOC state. 
The BTW model has been solved exactly using the commutative property of the
particle addition operator\cite{dd}. Several properties of this critical state,
e.g., entropy, height correlation, height probabilities, cluster statistics,
etc have been 
calculated analytically\cite{ddsnm,evi,vbp}.  And extensive numerical efforts
have also been performed to  estimate various critical 
exponents\cite{ssm1,grsm,ssm2,vbpevi,bb,lub,ev,aba1}. Recently, the avalanche 
exponents were estimated using the renormalization scheme\cite{piet,evrg}.
The BTW model in dilute lattice has also been studied recently\cite{aba2,najafi}.

It is important to mention that in the ricepile experiment,  Oslo observed that 
SOC occurs in systems of granular materials. And in the case of earthquakes, 
the power
law describing the magnitude of earthquake is similar to the power law followed 
in the case of frequency distribution of the size of avalanches in the sandpile 
model\cite{ Gros}.


So far in all the studies related to BTW model, the interaction with the
 nearest neighbouring sites are considered to be identical, i.e, 
the particles from the unstable lattice site, as it topples, 
distributed uniformly among its nearest neighbouring  sites. In case of 
BTW model on two dimensional regular lattice, the particles are distributed 
uniformly among the four nearest neighbouring sites from the unstable site.
However, in this study, we have considered that two particles are distributed 
uniformly along one direction (say along X-axis) i.e., one particle moves along
positive X-axis and the other moves along negative X-axis from the unstable
lattice site. But, along other (Y) direction, there is a probability($P_r$)
that the remaining two particles are distributed nonuniformly,
(i.e., both particles will go towards positive Y-axis  with probability
$P_r$). Consequently, two particles are distributed uniformy, with probability 
(1-$P_r$), restoring original BTW model.

In the original BTW model (with uniform distribution of particles after
topplings) in two dimensions, may be visualized as the sandpile formed on
a horizontal plane. Whereas, a probabilistically nonuniform distribution
of particle (considered in the present study), is a manifestation of
formation of sandpile on an inclined plane.

In this work, we have first studied the time evolution of the spatial average of
the height variable, $\bar z$, for a particular probability $P_r$, of nonuniform
distribution of particles along Y-axis.
The distribution of size and duration of the avalanches in the critical state
has also been obtained.
Here, we have also  studied the effect of probabilistically nonuniform
distribution of particles along Y-axis through the spatial varaition of 
the total number of topplings, $N_t(i,j)$, 
occurred at a particular lattice site ($i,j$),
during the  stipulated time of evolution of the system.

This paper is organised as follows. In Section I, the model and simulation
is discussed. In section II, the results are described. And the paper ends with
the conclusion, in Section III.

\bigskip

\leftline {\bf II. The model and simulation}

The BTW model is a lattice automata model of sandpile growth,
which evolves spontaneously into a critical state. 
We consider a two dimensional square lattice of size $L \times L$. 
Each site $(i,j)$ of the lattice is associated with a variable 
(so called height) $z(i,j)$ which can take positive integer values  varying 
from 0 to $z_c$.  In every time step, one particle
is added to a randomly selected site, which increases the value of the height
of that site, according to
\begin{equation}
z(i,j)=z(i,j)+1. 
\end{equation}

\noindent If, at any site the height variable exceeds a critical value
$z_c$ (i.e., if $z(i,j) \geq z_c$), then that site becomes unstable and
it relaxes by a toppling. When  an unstable site
topples, the value of the height variable of that site is reduced by 4
units and that of each of the four of its neighbouring sites  increased   
by unity (local conservation), i.e.,
\begin{equation}
z(i,j)=z(i,j)-4
\end{equation}
\begin{equation}
z(i \pm 1,j) = z(i\pm 1 ,j ) +1  ~~{\rm and}~~~~
z(i, j \pm 1) = z(i , j \pm 1) +1 
\end{equation}
\noindent for $z(i,j) \geq z_c$. Each boundary site is attached  to
an additional site which acts as a sink. We use here the 
open boundary conditions  so that the system can
dissipate  through the boundary. In our simulation, we have
taken $z_c = 4$. 

In original BTW model, when the unstable site topples,  four particles from the
 unstable site are distributed uniformly among its four nearest neighbours. 
In this paper, 
we consider the distribution of two particles from the unstable site equally
among the two nearest neighbouring sites  along the directions of X-axis.
 But in case of distribution of
particles along Y-axis, the distribution is probabilistically nonuniform. 
Thus, there is a probability($P_r$), that two particles from the unstable site 
will move  to the nearest neighbouring site along the positive direction 
of Y-axis, i.e., from $(i,j)$ to $(i,j+1)$. 
And with  probability ($1- P_r$),  two particles will be distributed equally
among the two nearest neighbouring sites  along both the directions of Y-axis.
Thus in this work, as the unstable site topples, the
particles from the unstable site will move towards its nearest neighbouring
 sites as follows:

\begin{equation}
z(i,j)=z(i,j)-4
\end{equation}
\begin{equation}
z(i \pm 1,j) = z(i\pm 1 ,j ) +1  ~~{\rm and}~~~~
\end{equation}
There is a probability ($P_r$) that two particles from the unstable site will
move along the direction of Y-axis  according to 
\begin{equation}
z(i, j + 1) = z(i , j + 1) +2 
\end{equation}
\begin{equation}
z(i, j - 1) = z(i , j - 1) 
\end{equation}
and with  probability ($1- P_r$), two particles   will  move  along  the two
direction of Y-axis according to
\begin{equation}
z(i, j \pm 1) = z(i , j \pm 1) +1 
\end{equation}

In this work, the system is evolved according to the  dynamics
(following eqns 4-8) starting from an initial condition with all the sites
having $z=0$. With the evolution of time, the value of height variables 
$(z(i,j)$ of different sites first increases due to random addition of 
particles. As soon as the value of height variable of any site reaches 
(or exceeds) the critical value ($z_c = 4$), that site topples.  
Here, we have studied the following observations in BTW model with
probabilistically nonuniform distribution of particles,

(1) The time evolution of the average (spatial)value of $z$,i.e.,
$$\bar z= (1/N)\sum_{i=1}^N z_i ~~~~~~~~~~~~~(N=L^2)$$

(2)The fraction of sites, $f_z$, having the height variable $z=0,1,2,3$, in the
critical state.

(3) The distribution of the  avalanche size($D(s)$), and avalanche time $\tau$
 ($D(\tau)$).

(4) The spatial variation of the total number of topplings occured 
at a site $(i,j)$ during total time $(t)$, $N_t(i,j)$.

\bigskip
\leftline {\bf III. Results}

In this paper, we have  studied the two dimensional BTW model,
when the  two particles are distributed uniformly  along X-axis.
But in case of distribution of particles along Y-axis,  there is a
probability, $P_r=0.4$, that two particles are distributed  nonuniformly 
(i.e, two particles move along the positive direction of Y-axis).
Here we have considered  a square lattice of size ($L = 200$).
We first studied  the time evolution  of the average (spatial)
value of $z$,  (i.e., $\bar z$), which is plotted in Figure-1(a).
Figure-1(a) shows that the value of $\bar z$ in critical state is 1.98.
It is to be noted that this value is less than the value observed,
2.12\cite{btw2} in case of BTW model.

We have also  studied how the value of $\bar z$ changes as the tendency of two 
particles to move along a particular direction of Y-axis increases.  We have 
calculated the value of $\bar z$ at critical state for different values of the 
probability of occuring nonuniform distribution of particles, ($P_r$).
In Figure-1(b), the variation of the value of $\bar z$ at critical state is
plotted for different value of $P_r$. 
The Figure-1(b) shows that the value of $\bar z$, decreases as $P_r$ increases.
And  the value of $\bar z$ becomes 1.7, when both the particles move along a 
particular direction of Y-axis from the unstable site, as it topples.
Various studies related to the structure of the lattice at the critical state,
like the fractal dimension and the fraction of the sites having different 
height variable has been studied in case of original BTW model\cite{grsm}.
Here, for BTW model with probabilistically nonuniform distribution, we have 
also calculated the fraction of sites($f_z$) for different height variables,
$z=0,1,2,3$. We have calculated $f_z$ for different probability, $P_r$ and 
plotted in Figure-2. The calculation of $f_z$ is done for a lattice of size,
L=200 and in the critical state reached at $t=4\times L^2$.

We have also calculated  the distributions of duration($\tau$) and size($s$) 
of the avalanches at the critical state. The distribution is obtained for 
80000 number of avalanches.
The distribution of avalanche size, $D(s)$ is plotted,
on a  doubly logarithmic scale, in Figure-3(a).
Similarly the  distribution of avalanche time, $D(\tau)$,
is plotted,  on a  doubly logarithmic scale,  in Figure-3(b). 
We have  estimated the value of the exponents within limited accuracy 
and given by,
$D(s) \propto s^{-1.22}$ and $D(\tau) \propto \tau^{-1.60}$.
The power law variations of the distribution of avalanche size ($s$) 
and avalanche time ($\tau$) given by:
$D(s)\propto s^{-1.22}$ and $D(\tau)\propto \tau^{-1.60}$, indicates
that the steady state is a critical state, however the exponents are different
from the BTW model.

Here we have calculated the total number of topplings, $N_t(i,j)$, occured at a 
site $(i,j)$ during the total time of evolution of the system and observed its 
spatial variation on the square lattice. 
In Figure-4(a), the value of total number of topplings occured at any site, 
$N_t(i,j)$, is plotted for different lattice sites  when the four particles
are distributed equally among the four nearest neighbouring sites from the
unstable site as it topples, i.e, for BTW model. 
Similarly, the Figure-4(b)  plots the spatial variation of $N_t(i,j)$
when there is a probability
($P_r=0.4$), that two particles will be nonuniformly distributed along 
the  Y-axis.
Here, we have plotted the spatial variation of the total number of topplings
occurred at a site, $N_t(i,j)$, for a square lattice of size, L=50.

It is interestingly observed that the spatial variation of the number of 
topplings, $N_t(i,j)$  is symmetric for the uniform distribution of particles 
from the unstable site. And it becomes asymmetric as the distribution of 
particles become probabilistically nonuniform along one direction from the 
unstable site as it topples. 

\newpage

\leftline {\bf IV. Summary}

We studied here, the BTW model with probabilistically nonuniform distribution
of two particles, from the unstable sites as it topples, among  its nearest 
neighbouring sites along a particular 
direction. It is observed that in the case of nonuniform distribution, the
spatial average value of the height variable, $\bar z$, reaches a steady value,
which is less than the value that obtained in the BTW model with uniform 
distribution of particles. The exponents of the power law obeyed by the
distribution of  size($s$) and duration($\tau$) of the avalanches 
are calculated as $D(s)\propto s^{-1.22}$ and $D(\tau)\propto \tau^{-1.55}$.
The fractions of the lattice sites, $f_z$, having different height variables, 
$z=0,1,2,3$, at the critical state have been calculated. And the variations
of $f_z$ with the prbabiliity $P_r$ have been studied.
The total number of topplings, $N_t(i,j)$, occured at any site, $(i,j)$,
 is calculated and plotted for different lattice sites for both the cases, 
uniform and probabilistically nonuniform distribution of particles among 
the four nearest neighbouring sites from the unstable site. 
The total number of topplings occurred at the central site is maximum and 
it is symmetric with respect to X and Y axis in case of uniform distribution
of particles.  But in case of probabilistically nonuniform distribution of 
particles from the unstable site  the symmetry observed in the spatial variation
of the quantity, total number of topplings occurred at any site, is broken.

\newpage
\setlength{\unitlength}{0.240900pt}
\ifx\plotpoint\undefined\newsavebox{\plotpoint}\fi
\sbox{\plotpoint}{\rule[-0.200pt]{0.400pt}{0.400pt}}%


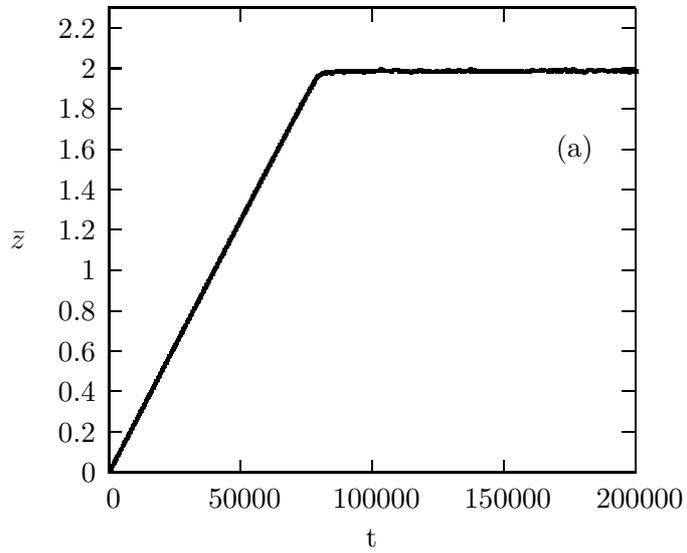
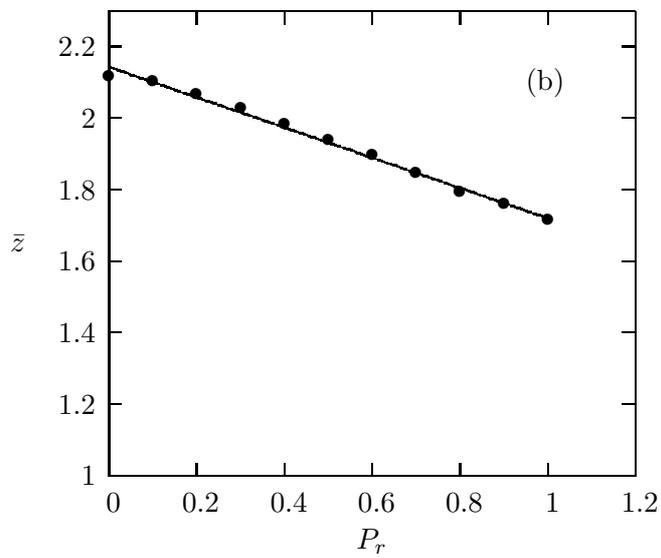
\noindent {\bf Fig-1.} (a)Plot of time variation of the spatial average of 
height variable, ($\bar z$) for probabilistically nonuniform distribution
of particles along Y-axis from the unstable site ($P_r=.4$).
(b)Plot of  $\bar z$ against $P_r$. (L=200)

\newpage
\setlength{\unitlength}{0.240900pt}
\ifx\plotpoint\undefined\newsavebox{\plotpoint}\fi
\sbox{\plotpoint}{\rule[-0.200pt]{0.400pt}{0.400pt}}%
%

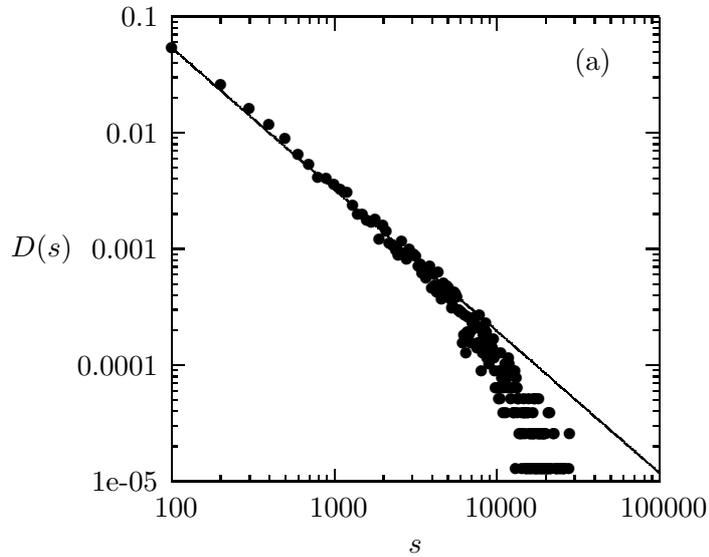
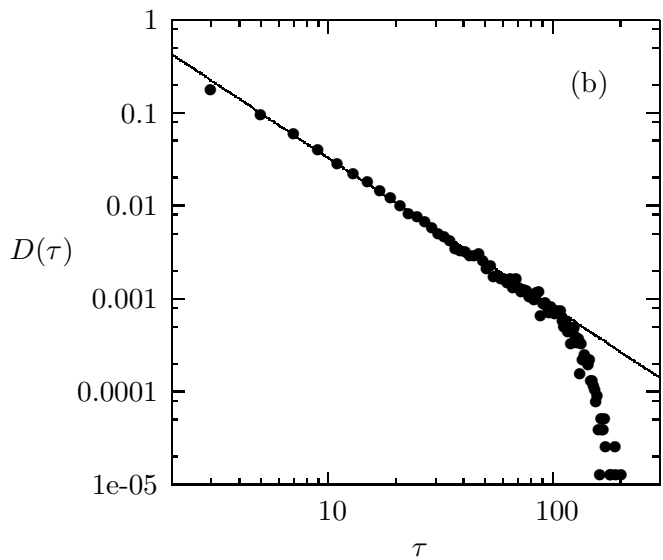
\noindent {\bf Fig-3.} (a)Log-log plot of  distribution of avalanche size
 ($s$) for probabilistically nonuniform distribution of particles along Y-axis 
for  a particular probability, $P_r=0.4$.
Solid line represents $y\sim x^{-1.22}$.
(b)Log-log plot of  distribution of avalanche time ($\tau$)
for nonuniform distribution of particles along Y-axis for   a particular
probability, $P_r=0.4$.
Solid line represents $y\sim x^{-1.60}$.
\newpage

\setlength{\unitlength}{0.240900pt}
\ifx\plotpoint\undefined\newsavebox{\plotpoint}\fi
\sbox{\plotpoint}{\rule[-0.200pt]{0.400pt}{0.400pt}}%


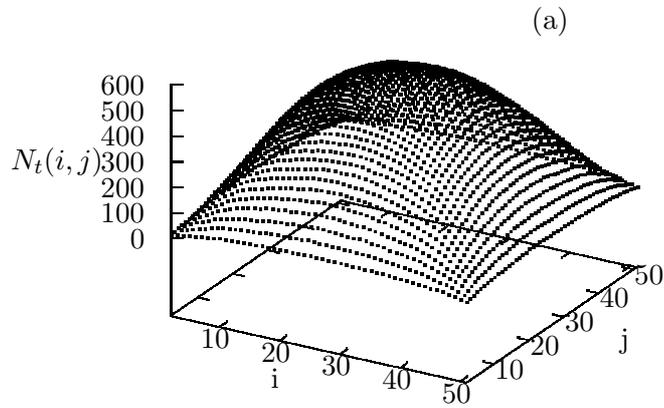
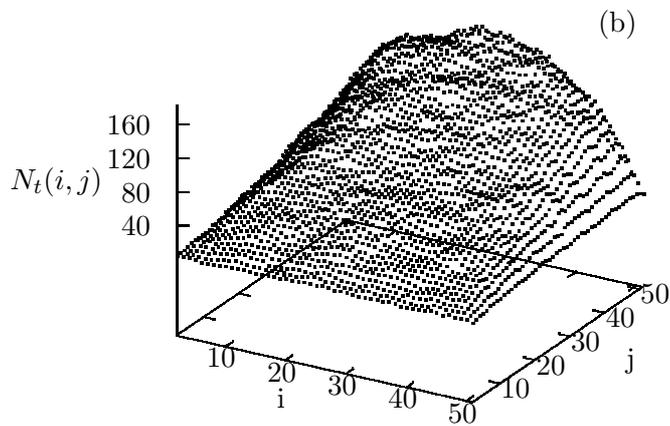
\noindent {\bf Fig-4.} Plots the variation of $N_t(i,j)$ for different sites of
 the lattice for (a) uniform distribution of the particles among the nearest
 neighbouring sites and (b) for probabilistically nonuniform distribution
of particles with a particular probability, $P_r=.4$.
\newpage

\begin{center}{\bf References}\end{center}
\begin{enumerate}

\bibitem{btw1} P. Bak, C. Tang and K. Wiesenfeld, Phys. Rev. Lett., {\bf 59}
(1987) 381

\bibitem{btw2} P. Bak, C. Tang and K. Wiesenfeld, Phys. Rev. A, {\bf 38} (1988) 364.

\bibitem{dd} D. Dhar, Phys. Rev. Lett., {\bf 64} (1990) 1613;

\bibitem{ddsnm} S. N. Majumder and D. Dhar, J. Phys. A {\bf 24} (1991) L357.

\bibitem{evi} E. V. Ivashkevich, J. Phys. A {\bf 27} (1994) 3643;

\bibitem{vbp} V. B. Priezzhev, J. Stat. Phys, {\bf 74} (1994) 955.

\bibitem{ssm1} S. S. Manna, J. Stat. Phys. {\bf 59} (1990) 509

\bibitem{grsm} P. Grassberger and S. S. Manna, J. De. Physique. {\bf 51} (1990) 1077;

\bibitem{ssm2} S. S. Manna, Physica A {\bf 179} (1991) 249

\bibitem{vbpevi} V. B. Priezzhev, D. V. Ktitarev and E. V. Ivashkevich,
Phys. Rev. Lett {\bf 76} (1996) 2093.

\bibitem{bb} A. Benhur and O. Biham, Phys. Rev. E {\bf 53} (1996) R1317.

\bibitem{lub} S. Lubek and K. D. Usadel, Phys. Rev. E {\bf 55} (1997) 4095.

\bibitem{ev} E. V. Ivashkevich, J. Phys. A {\bf 27 } (1994) L585.

\bibitem{piet} L. Pietronero, A. Vespignani and S. Zapperi, Phys. Rev. Lett.,
{\bf 72} (1994) 1690.

\bibitem{evrg} E. V. Ivashkevich, Phys. Rev. Lett, {\bf 76} (1996) 3368.

\bibitem{aba1} A. B. Acharyya, Acta Phys. Pol. B, {\bf 42}, (2011) 19.

\bibitem{aba2} A. B. Acharyya, J. Mod. Phys, {\bf 5}, (2014) 1958.


\bibitem{najafi} M. N. Najafi, J. Phys. A {\bf 49 } (2016) 335003.

\bibitem{Gros} Claudius Gros, {\it Complex and Adaptive Dynamical Systems}, Springer, 2008.




\end{enumerate}
\end{document}